# Ni-Cr textured substrates with reduced ferromagnetism for coated conductor applications

J. R. Thompson,[a,b] A. Goyal,[a] D. K. Christen,[a] and D. M. Kroeger[a]

[a]Oak Ridge National Laboratory, PO Box 2008, Oak Ridge TN 37831-6061 USA
[b]Department of Physics, University of Tennessee, Knoxville TN 37996-1200

**Abstract**  File: ReducedFMsubstrates4.doc

A series of biaxially textured $Ni_{1-x}Cr_x$ materials, with compositions $x = 0, 7, 9, 11$, and 13 at % Cr, have been studied for use as substrate materials in coated conductor applications with high temperature superconductors. The magnetic properties were investigated, including the hysteretic loss in a Ni-7 at.% Cr sample that was controllably deformed; for comparison, the loss was also measured in a similarly deformed pure Ni substrate. Complementary X-ray diffraction studies show that thermo-mechanical processing produces nearly complete {100}<100> cube texturing, as desired for applications.

**Introduction**

A major application of high temperature superconductors (HTS) is the development of tapes and cables for low loss conduction of high density electric currents. First to be developed were conductors incorporating Bi-cuprates, which are more easily textured to minimize the vexing problem of "weak-link" intergrain current transport. These first generation conductors suffer, however, from poor performance at elevated temperatures in large magnetic fields, together with economic concerns stemming from the usage of silver as the preferred cladding and substrate material. Subsequently, second generation coated conductors have been developed, based largely on YBaCuO superconductors. Three major approaches have been developed: the "RABiTS" (*R*olling *A*ssisted *Bi*axially *T*extured *S*ubstrates) method [1-3], Ion Beam Assisted Deposition (IBAD) [4-6], and the Inclined Substrate Deposition (ISD) method [7,8]. In each case, a major objective is to create a substrate on which highly textured HTS with low angle grain boundaries can conduct large intergrain currents. IBAD and ISD methods form a textured buffer layer on an untextured polycrystalline metallic tape for subsequent epitaxial deposition of c-axis textured YBaCuO superconductor. The RABiTS process achieves the critically important grain alignment by formation of a thermomechanically produced biaxially textured metallic substrate, which is subsequently coated with appropriate buffer layers followed by epitaxial deposition of the c-axis textured superconductor.

The first metals used for biaxial texturing for superconductive applications were Ag and Ni [1]. Silver is technically difficult, however, due to its lack of strength, cost, and difficulty of obtaining a sharp cube texture [1]. Nickel lends itself well for the desired {100}<100> texture, and most of the work to date has used pure Ni as a base. An obvious consideration, however, is its ferromagnetism with a Curie temperature of 627 K and a saturation magnetization of 57.5 emu/g at $T = 0$. The ferromagnetism likely complicates the design of high field magnets for critical applications (magnetic resonance imaging [MRI], accelerators, etc.). Furthermore, usage of Ni-based tapes in alternating current (ac) applications runs the risk of increased energy loss, due to hysteretic loss in the magnetic material. The actual magnetic energy loss per cycle of applied magnetic field $H$ is given by the area enclosed within a magnetization loop $M(H)$, and this varies with the magnetic "hardness" of the ferromagnet.

With these considerations, it is obviously desirable to develop suitable *alloys with reduced ferromagnetism* (FM) that can also be successfully biaxially textured. The difficulty in forming such substrates stems from the fact that even minor additions of alloying additions greatly affect the ability to produce a dominant biaxial texture in pure FCC metals. Recently, a method has been proposed to fabricate cube textured, alloy substrates with reduced magnetism [9]. In this work, we report on the magnetic properties of a series of Ni-Cr alloys possessing lower Curie temperatures $T_c$ and lower saturation magnetization $M_{sat}$. As demonstrated below, these materials reduce or eliminate the potentially undesirable FM of pure Ni, while retaining the ability to be biaxially textured.

**Experimental aspects**

A series of alloys was prepared, using 99.99 % purity starting elements. Appropriate mixtures were vacuum arc melted and dropped into copper moulds to form rods of the appropriate composition. We investigated $Ni_{1-x}Cr_x$ materials with nominal compositions $x = 0, 7, 9, 11,$ and 13 at % Cr. Details of composition, obtained by inductively coupled plasma (ICP) analysis, are given in Table I. The materials were biaxially textured by rolling deformation followed by annealing at 1050 $^o$C for 2 hours in vacuum with an oxygen partial pressure $< 10^{-7}$ Torr. For magnetization measurements, the sheets of thickness 0.13 mm were cut into 5 mm x 5 mm pieces. Generally, a stack of 3-5 pieces was mounted in a thin plastic tube for magnetic measurements with the magnetic field $H$ applied parallel to the plane of the sheets, to minimize demagnetizing effects. In an additional experiment to assess the magnitude of energy loss due to ferromagnetic hysteresis after a representative amount of work-hardening, a foil of $Ni_{93}Cr_7$ was controllably deformed. The foil was wrapped around a 9 mm mandrel, flattened, then reverse-wrapped, and again flattened for one deformation cycle. It was studied at 77 K with field $H$ applied either parallel or perpendicular to the plane of the foil. For comparison, the hysteretic loss in biaxially textured Ni, after 3 complete deformation cycles, was investigated as well.

The magnetic studies were conducted in a SQUID-based magnetometer at temperatures $T = $ 5 - 300 K, in fields $H$ up to 65 kOe. We measured both the isothermal mass magnetization $M(H)$ at various temperatures and the "susceptibility" $M/H$ versus $T$ in fixed field, normally with $H = 1$ kOe. Dimensionally, the mass magnetization $M$ = (magnetic moment)/(mass of alloy) has units of emu/gram = G-cm$^3$/g. To record powder diffraction patterns, a Philips model XRG3100 diffractometer with Cu K$_\alpha$ radiation was used. A Picker four-circle diffractometer was used to determine the texture of the films by omega and phi scans. Pole figures were collected to determine the percentage of cube texture.

**Magnetic Properties of Textured NiCr alloys**

To provide an overview of the magnetic properties of the $Ni_{1-x}Cr_x$ alloys, Fig. 1 shows the temperature dependence of the mass magnetization $M(T)$, measured in an applied magnetic field $H$

= 1 kOe. Qualitatively, it is evident that $M$ decreases quickly with the addition of Cr. Also, the Curie temperature $T_c$, noted on the figure, steadily diminishes. The $T_c$ values were obtained using the relation that the spontaneous magnetization $M \propto (T_c - T)^\beta$ with $\beta = 1/3$.[10] Figure 2 illustrates this dependence by plotting $M^3$ versus $T$, wherein a linear extrapolation to $M^3 = 0$ yields values for $T_c$. In this process, we ignore data very close to $T_c$, where $M$ is influenced by presence of the applied field. The power law dependence describes the data rather well for a substantial temperature range below $T_c$. Results for the Curie temperature are shown in Fig. 1 and listed in Table I for all alloys.

For coated conductor applications, one anticipates that the materials will be employed in the presence of large magnetic fields and for generating such fields. Thus it is useful to examine the field dependent magnetization, which is shown in Fig. 3 at representative temperatures of 40 and 77 K. At these temperatures, it is evident from the "square" response of the 7 and 9 at % Cr alloys that they are ferromagnetic. In contrast, the alloys with higher Cr-content are paramagnetic and have significantly lower magnetization. From similar studies at 5 K, we obtained the saturation magnetization $M_{sat}$ values listed in Table I. Included in Fig. 3 are data for annealed Type 304 stainless steel, a common construction material in cryogenic applications; a comparison shows that the magnetization of the Ni-11 at.% Cr alloy is comparable with the 304 SS and the 13 at. % Cr alloy is even lower.

Let us now consider the paramagnetic state above $T_c$. For an applied field that is not too large (here $H = 1$ kOe), the ratio $M/H$ closely approximates the initial differential susceptibility $\chi_{mass} = dM/dH$. Thus we analyze the data using a Curie-Weiss dependence

$$M/H = C/(T - \theta_P) \tag{1}$$

where $C = N_A \mu_B^2 p_{eff}^2 / 3k_B$ is the Curie constant and $\theta_p$ is the paramagnetic Weiss temperature. As usual, $N_A$ is Avogadro's number, $\mu_B$ is the Bohr magneton, $p_{eff}$ is the effective magnetic moment per atom, and $k_B$ is Boltzmann's constant. Thus we plot the reciprocal $H/M$ versus $T$ in Fig. 4, which shows a reasonable description of the data. The resulting values for the Weiss temperatures are shown in Table I. At higher temperatures, some samples deviate from the simple Curie-Weiss relationship; this may be due to deviations from the overly simple temperature dependence of Eq. 1 or the presence of additional paramagnetic contribution(s), e.g., nominally temperature-independent orbital terms not included in this expression. As is frequently the case for ferromagnetic materials, the values for $\theta_p$ lie near the respective Curie temperatures, but rarely coincide exactly. In Table I, along with the paramagnetic Weiss temperatures $\theta_p$, are values of the temperature at which $d(M/H)/dT$ is largest; the latter provides a rough estimate of the Curie temperature, provided that the applied field (here 1 kOe) is not too large.

Some qualitative dependencies on Cr content $x$ are already apparent. More quantitative features are presented in Fig. 5. In Fig. 5a is the Curie temperature $T_c$ that decreases quite linearly with $x$, as shown by the solid line fitted to the present data. A simple linear extrapolation to the axis at $T_c = 0$ intersects at critical Cr concentration of $11.5 \pm 0.4$ at %. This value lies slightly below the critical concentration $x_c = 13$ at % Cr deduced previously by Besnus et al.[11] These authors, whose results are included in Fig. 5, observed a "tail" in the $T_c(x)$ dependence near $x_c$. Included, too, are earlier results from the compendium of Bozorth [12], in which many values were obtained by extrapolating from higher temperatures. The present results for our biaxially textured substrate materials lie close to those from some earlier studies [10, 13, 14], but significantly below the values reported in Bozorth.[12] The second frame, Fig. 5b, presents the saturation magnetization $M_{sat}$ at $T = 5$ K. Again a linear decrease with Cr-content is observed, as illustrated by the solid line, with an intersection at a Cr-concentration of $12.0 \pm 0.2$ at %. As in Fig. 5a, the present results lie below the earlier data from Bozorth and agree within experimental error with the data of Besnus et al. [10]

Next we consider the hysteretic energy loss of two controllably deformed, ferromagnetic substrate materials, the Ni-7 at % Cr alloy and for comparison, a biaxially textured strip of pure Ni. For the alloy, a foil of Ni-7 at % Cr was wrapped around a 9 mm mandrel, flattened, reverse wrapped, then again flattened to simulate the work hardening that might be encountered in fabricating a superconducting component. The magnetization of the material is shown in Fig. 6 with the field applied either parallel to plane of the foil (steep curves related to a small demagnetizing factor) or parallel to a normal **n** to the surface (flatter curve with much stronger demagnetizing effects). In either case, the material is relatively reversible, qualitatively indicating a modest energy loss due to magnetic hysteresis. Numerical integration of the loop area gives a hysteretic energy loss/cycle of 160 erg/g = 1400 erg/cm$^3$ with H $\parallel$ **n** and 230 erg/g = 2000 erg/cm$^3$ with H $\perp$ **n**. The coercive field at 77 K is 4 Oe.

The corresponding hysteretic loss was measured in an unalloyed tape of pure biaxially textured Ni. A 4 x 5 mm$^2$ foil was subjected to three complete cycles of deformation, also around a 9 mm mandrel. Its magnetization loop was measured at 77 K in the range $\pm 800$ Oe (we limited the field to this interval to avoid trapping flux in the superconductive magnet, which would distort the measurement.) The results are shown in Fig. 7. The steeper curve, with H $\perp$ **n**, approaches saturation. Integrating the area within this loop gives an energy loss/cycle of 2050 erg/gm = 18200 erg/cm$^3$, and a coercive field of 7 Oe. The flatter curve with H $\parallel$ **n** does not approach saturation, due to the large demagnetizing factor in this orientation and the higher saturation magnetization of pure Ni, 57.5 erg/gm = 512 G at $T = 0$. With this minor hysteresis loop, the area is notably lower, giving a loss/cycle of 500 erg/gm = 4450 erg/cm$^3$. The inset in Fig. 7 shows the magnetization in applied fields large enough to saturate the material.

It is informative to compare the hysteretic losses in these substrates with the hysteretic loss arising from the superconductive material per se. For an estimate, we assume a typical RABiTS tape architecture with an 8 mm wide tape coated with 2.5 μm of YBCO, deposited on a 50 μm thick metal substrate. With a critical current density $J_c = 1 \times 10^6$ A/cm$^2$ at 77 K, this gives a critical current $I_c = 200$ A; for ac operation at 60 Hz, we assume a peak ac current $I_0 = I_c/2 = 100$ A. To estimate the superconductive energy loss, we use the theory of Norris [15], which applies to a long isolated conductor with a $J_c$ that is independent of field. This theory, for the case of a round or elliptic wire, has been shown to provide a good description of the ac loss in YBCO coated conductors on Ni [16] and on Hastalloy [17, Suenaga?]. For the extreme case that $I_0 = I_c$, the Norris expression gives a loss per cycle per m (SI units) of $L_c = (1/2\pi)\mu_0 I_0^2 = 8$ mJ/m-cycle; for $I_0 = I_c/2$, the loss is smaller by a factor of ~17 according to Table 3 of Norris [15], giving $L_c = 0.46$ mJ/m-cycle. To obtain the common figure-of-merit for a coated conductor, the power loss per kiloampere-meter of composite material, one multiplies by the ac frequency, 60 Hz, and the current ratio (1000 A/$I_0$) = 10. This yields a superconductive component to the hysteretic loss of 0.27 W/kA-m at $I_c/2$ (and 2.4 W/kA-m at $I_c$).

To estimate the additional loss due to the ferromagnetic substrate, note that in this example, there is 0.4 cm$^3$ of alloy per meter of tape. Assuming that the ambient ac field is sufficiently strong to carry the substrate around the entire hysteretic part of its $M(H)$ loop, one has a ferromagnetic loss per cycle per m of $L_c$ = 0.07 mJ/m for the 7 at % Cr alloy, and 0.7 mJ/m for the pure Ni substrate. The corresponding power losses are 0.043 W/kA-m and 0.43 W/kA-m, respectively. Thus the alloy substrate increases the loss by ~ 16 %, while the loss from the Ni substrate is comparable with that in the superconductor.

This example illustrates the magnitude of losses that might be encountered in power line applications. Note, however, that for currents $I_0$ near $I_c$, the superconductive energy loss $L_c$ increases rapidly with $I_0$, faster than $I_0^3$, while it varies as $(I_0/I_c)^3$ when the current is small. Since the superconductive and ferromagnetic losses have different dependencies on current $I_0$, their relative magnitude depends on how near to $I_c$ the coated conductor is driven. In addition, the above illustration assumes the worst case, namely that the ferromagnetic loss is maximized: in a conductor and field configuration where the substrate experiences smaller ac field amplitudes, the ferromagnetic loss component diminishes. This decrease is evident in the lower loss per cycle for the Ni material with **H** applied normal to the surface, where $H < 800$ Oe. Thus the total loss will depend not only on the choice of substrate materials, but also on the operating conditions $I_0/I_c$ and the conductor layout that affects the ac field experienced by the conductor. All of these losses, however, are smaller than the loss of ~ 1 W/(kA-m) typically cited for Ag-clad, BiSrCaCuO-based first generation tapes.

Overall, it appears that 7 at. % Cr alloy should be entirely satisfactory for many applications, obviating the need for higher Cr contents, for which the biaxial texturing is more difficult. The hysteretic losses in the deformed pure Ni are more significant and their impact must be evaluated in light of the intended application and the ac field environment.

**Texture in Ni-Cr alloys**

All alloy compositions studied were successfully textured to obtain ~100% cube texture. In Fig. 8, this is illustrated by the background corrected, (111) pole figure for a Ni-13 at. % Cr substrate annealed at 1050 °C for 2 hours. Only four crystallographically equivalent peaks, corresponding to the {100}<100> cube orientation, are present in the pole figure. Quantification of the pole figure suggests ~ 100 % cube texture in the sample. Figure 9 shows a φ-scan revealing the degree of in-plane texturing in the substrate. The full-width-half-maximum (FWHM) as determined from the φ-scan is 7.8 °. Figure 10 shows ω-scans to demonstrate the degree of out-of-plane texturing in the substrate. Figure 10a was obtained with the sample rotated in the rolling direction. The FWHM of the texture is 5.7 °. In Fig. 10b is an omega scan with the sample rotated about the rolling direction, for which the FWHM is 8.9 °. Clearly, these results demonstrate that Ni-Cr alloys with Cr contents as high as 13 at % Cr can be thermo-mechanically processed to form substrates with very sharp and fully formed cube texture.

**Conclusions**

This work shows that Ni-Cr alloys with much reduced or non-existent ferromagnetism can be fabricated and successfully processed to have biaxial cube texturing. The hysteretic loss in a deformed sample of alloy was measured and compared with that in pure Ni; the results suggest that the magnetic loss of the 7 at % Cr substrate in a coated conductor should be small compared with that in first generation Bi-cuprate tapes and small compared with hysteretic losses in the superconductive coating.

We wish to thank K. D. Sorge for experimental assistance; H. R. Kerchner for scientific discussions; and R. E. Ericson and C. V. Hamilton of the 3M company, St. Paul, MN for chemical analysis of the alloys. Research was co-sponsored by the DOE Division of Materials Sciences and the DOE Office of Energy Efficiency and Renewable Energy, Power Technologies, under contract DE-AC05-00OR22725 with the Oak Ridge National Laboratory, managed by UT-Battelle, LLC.

**Figure captions**

Fig. 1. The mass magnetization of $Ni_{1-x}Cr_x$ alloys versus temperature, measured in an applied field $H = 1$ kOe applied parallel to the plane of the foils. Nominal Cr concentration $x$ is given in atomic %.

Fig. 2. A plot of $M^3$ versus temperature $T$. Straight lines show the extrapolation to $M = 0$ used to define the Curie temperature $T_c$.

Fig. 3. The field dependence of the magnetization of $Ni_{1-x}Cr_x$ materials at 40 K (closed symbols) and at 77 K (open symbols). For comparison, the magnetization of annealed Type 304 stainless steel at 77 K is shown as a dotted line.

Fig. 4. The inverse "susceptibility" $H/M$ versus temperature $T$ for three ferromagnetic NiCr alloys. The straight lines show a simple Curie-Weiss dependence (Eq. 1) for alloys with 9 and 11 at % Cr.

Fig. 5. Variation of magnetic properties of $Ni_{1-x}Cr_x$ alloys with Cr content $x$. (a) the Curie temperatures $T_c$ obtained here and those reported by Bozorth [12] and by Besnus et al. [11]. The line is a linear dependence fitted to the present $T_c(x)$ data. (b) The saturation magnetization at $T = 5$ K versus $x$.

Fig. 6. Magnetization loops (expanded scale) for a deformed $Ni_{93}Cr_7$ foil at 77 K, with magnetic field applied parallel or perpendicular to the plane of the foil. The magnetization is relatively reversible, with limited hysteretic energy loss/cycle.

Fig. 7. Magnetization loops (expanded scale) for a deformed biaxially textured Ni foil at 77 K, with magnetic field applied parallel or perpendicular to the plane of the foil. Inset: magnetization in large fields.

Fig. 8. A (111) pole figure for a Ni-13 at% Cr substrate that was annealed at 1050 °C for 3 hours.

Fig. 9. A φ-scan showing the in-plane texturing in the substrate (same as Fig. 7). The full-width-half-maximum (FWHM) is 7.8°.

Fig. 10. Angular ω-scans showing the degree of out-of-plane texturing (same substrate as Fig. 7). (a) sample rotated in the rolling direction, giving a FWHM of the texture of 5.7 °; (b) sample rotated about the rolling direction, with FWHM of 8.9 °.

**References**
[1] A. Goyal, D. P. Norton, J. D. Budai, M. Paranthaman, E. D. Specht, D. M. Kroeger, D. K. Christen, Q. He, B. Saffian, F. A. List, D. F. Lee, P. M. Martin, C. E. Klabunde, E. Hatfield and V. K. Sikka, Appl. Phys. Lett. **69,** 1795 (1996) .
[2] A. Goyal , J. Budai, D. M. Kroeger, D. Norton, E. D. Specht and D. K. Christen, US Patents: 5, 739, 086; 5, 741, 377; 5, 898, 020; 5, 958, 599.
[3] A. Goyal, R. Feenstra, F. A. List, M. Paranthaman, D. F. Lee, D. M. Kroeger, D. B. Beach, J. S. Morrell, T. G. Chirayil, D. T. Verebelyi, X. Cui, E. D. Specht, D. K. Christen and P. M. Martin, J. of Metals **51**, 19 (1999).
[4] Y. Iijima, N. Tanabe, O. Kohno, and Y. Ikeno, Appl. Phys. Lett. **60**, 769 (1992); Y. Iijima, K. Onabe, N. Futaki, N. Sadakata and O. Kohno, J. Appl. Phys. Lett. **74**, 1905 (1993).
[5] R. P. Reade, P. Burdahl, R. E. Russo and S. M. Garrison, Appl. Phys. Lett. **61**, 2231 (1993).
[6] X. D. Wu, S. R. Foltyn, P. N. Arendt, W. R. Blumenthal, I. H. Campbell, J. D. Cotton, J. Y. Coulter, W. L. Hults, M. P. Maley, H. F. Safar and J. L. Smith,  Appl. Phys. Lett. **67**, 2397 (1995).
[7] K. Hasegawa, H. Fujino, H. Mukai, M. Konishi, K. Hayashi, K. Sato, S. Honjo, Y. Sato, H. Ishii and Y. Iwata, Appl. Supercond. **4**, 487-496 (1996).
[8] M. Fukutomi, S. Aoki, K. Kimori, R. Chatterjee, K. Togano and H., Maeda, Physica  C **231**, 113 (1994).
[9] A. Goyal, E. D. Specht, D. M. Kroeger, M. Paranthaman, US Patent 5964966
[10] *Solid State Physics*, N. W. Ashcroft and N. D. Mermin (Holt, Rinehart and Winston, New York, 1976), p. 699.
[11] M. J. Besnus, Y. Gottehrer, and G. Munshy, Phys. Stat. Sol. B **49**, 597 (1972).
[12} *Ferromagnetism* by R. P. Bozorth (IEEE Press, Piscataway, NJ, 1978), pp . 307-8.
[13] V. Suresh Babu, A. S. Pavlovic, and Mohindar S. Seehra, J. Appl. Phys. **79**, 5230 (1996).
[14] R. Chiffey and T. J. Hicks, Phys. Lett. **34A**, 267 (1971).
[15] W. T. Norris, J. Phys. D **3**, 489 (1970).
[16] H. R. Kerchner, D. P. Norton, A. Goyal, J. D. Budai, D. K. Christen, D. M. Kroeger, E. D. Specht, Q. He, M. Paranthaman, D. F. Lee, B. C. Sales, F. A. List, and R. Feenstra, Appl. Phys. Lett. **71**, 2029 (1997).
[17] M. Ciszek, O. Tsukamoto, N. Amemiya, J. Ogawa, O. Kasuu, H. Ii, K. Takeda K, and M. Shibuya, IEEE Trans. on Appl. Supercond. **10**, 1138 (2000).

| Nom comp. | Cr $x$ (at %) | $T$ at peak $(d\chi/dT)$ (K) | $T_c$ (K) | $\theta_p$ (K) | $M_{sat}$ (G-cm$^3$/g) |
|---|---|---|---|---|---|
| pure Ni | 0 | | 627 | -- | 57.5 |
| Cr-7 % | 7.2 | 245 | 248 | ~260 | 23.1 |
| Cr-9 % | 9.2 | 120 | 115 | 155 | 12.6 |
| Cr-11 % | 11.1 | 17 | 20 | 37 | 4 |
| Cr-13 % | 13.2 | -- | <10 | -- | 0.4 |

**Table I** Magnetic Properties of Ni-Cr Alloys

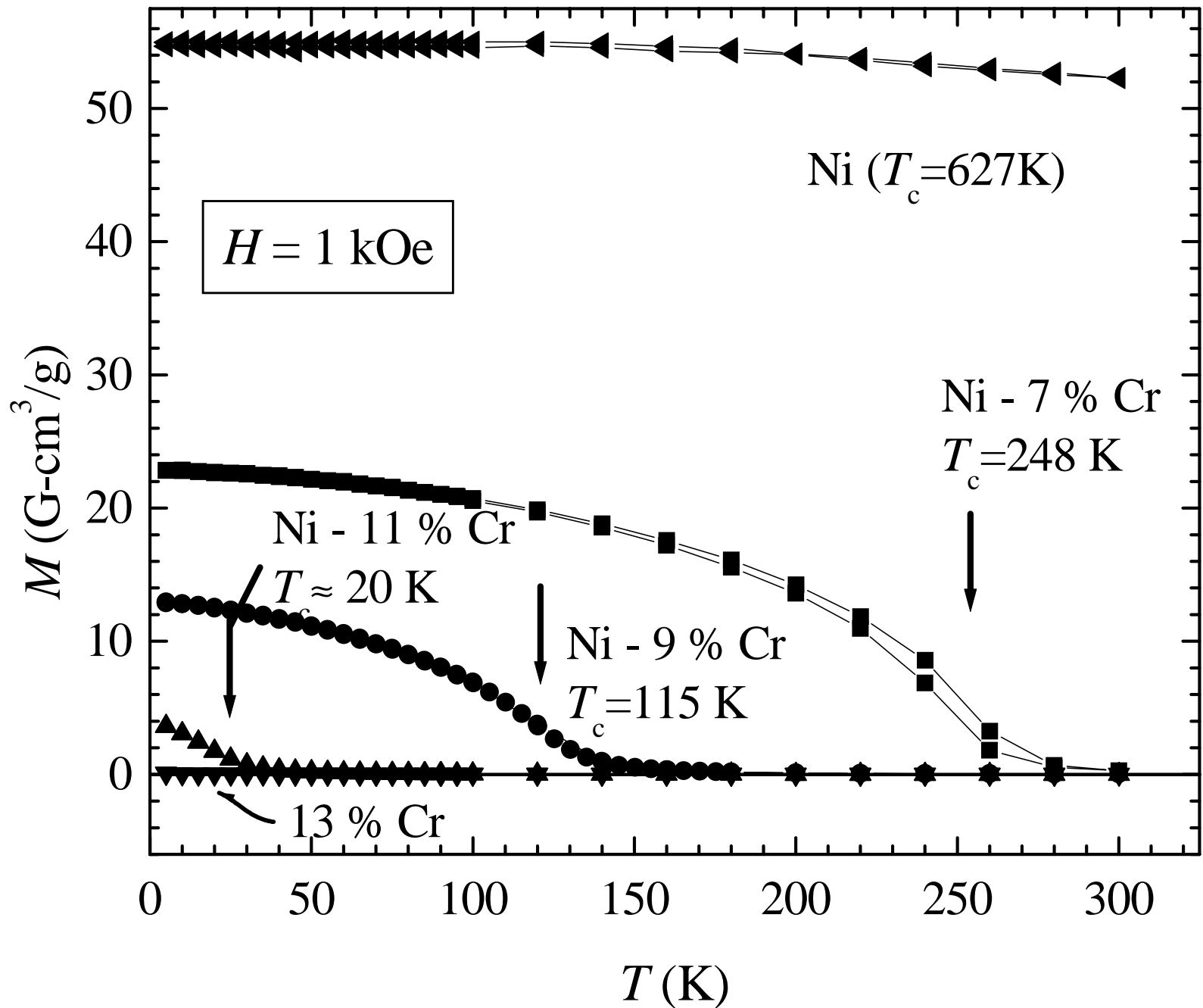

Fig. 1 JR Thompson et al.  C data NiCr

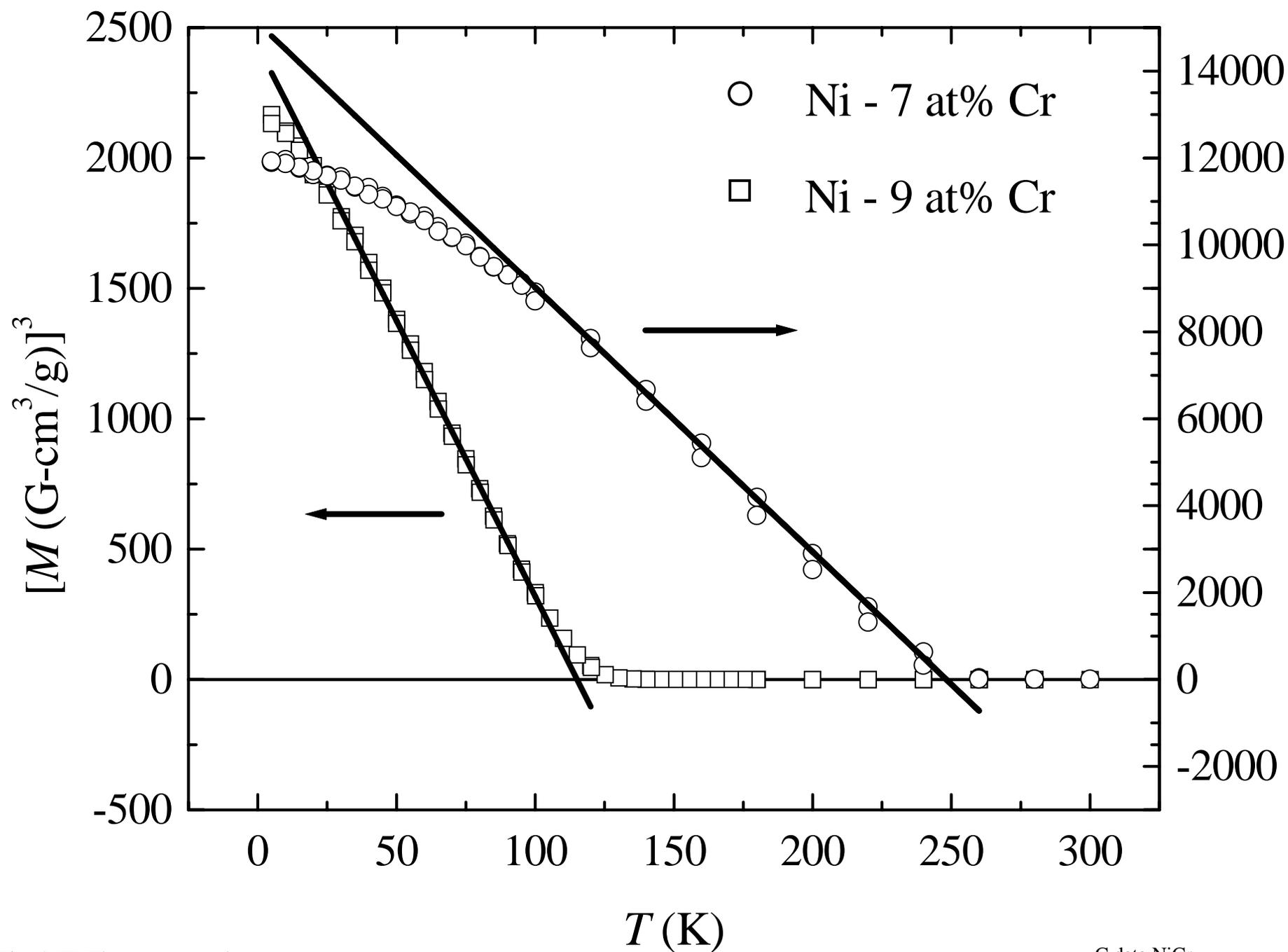

Fig. 2 JR Thompson et al.

C data NiCr

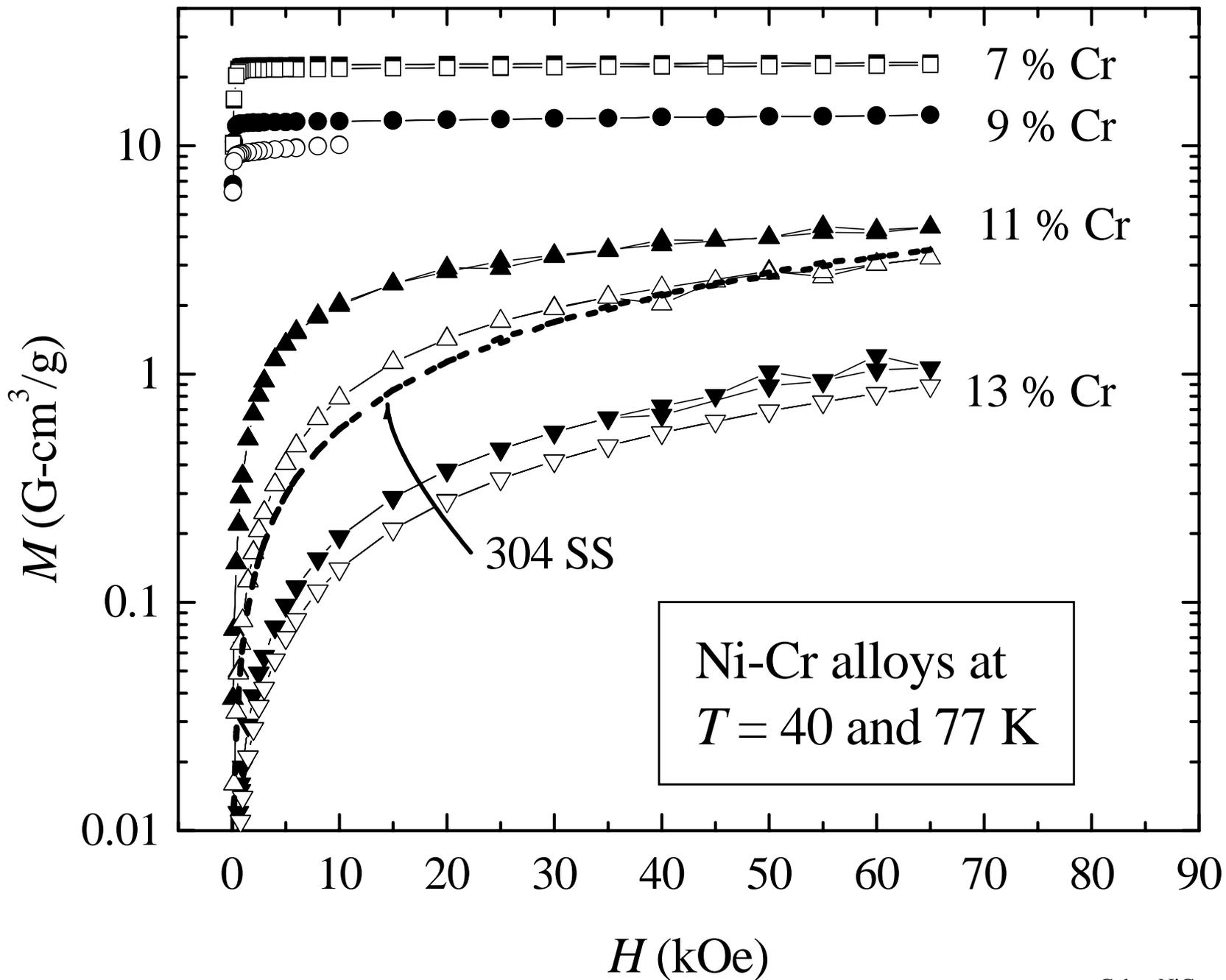

Fig. 3 JR Thompson et al.

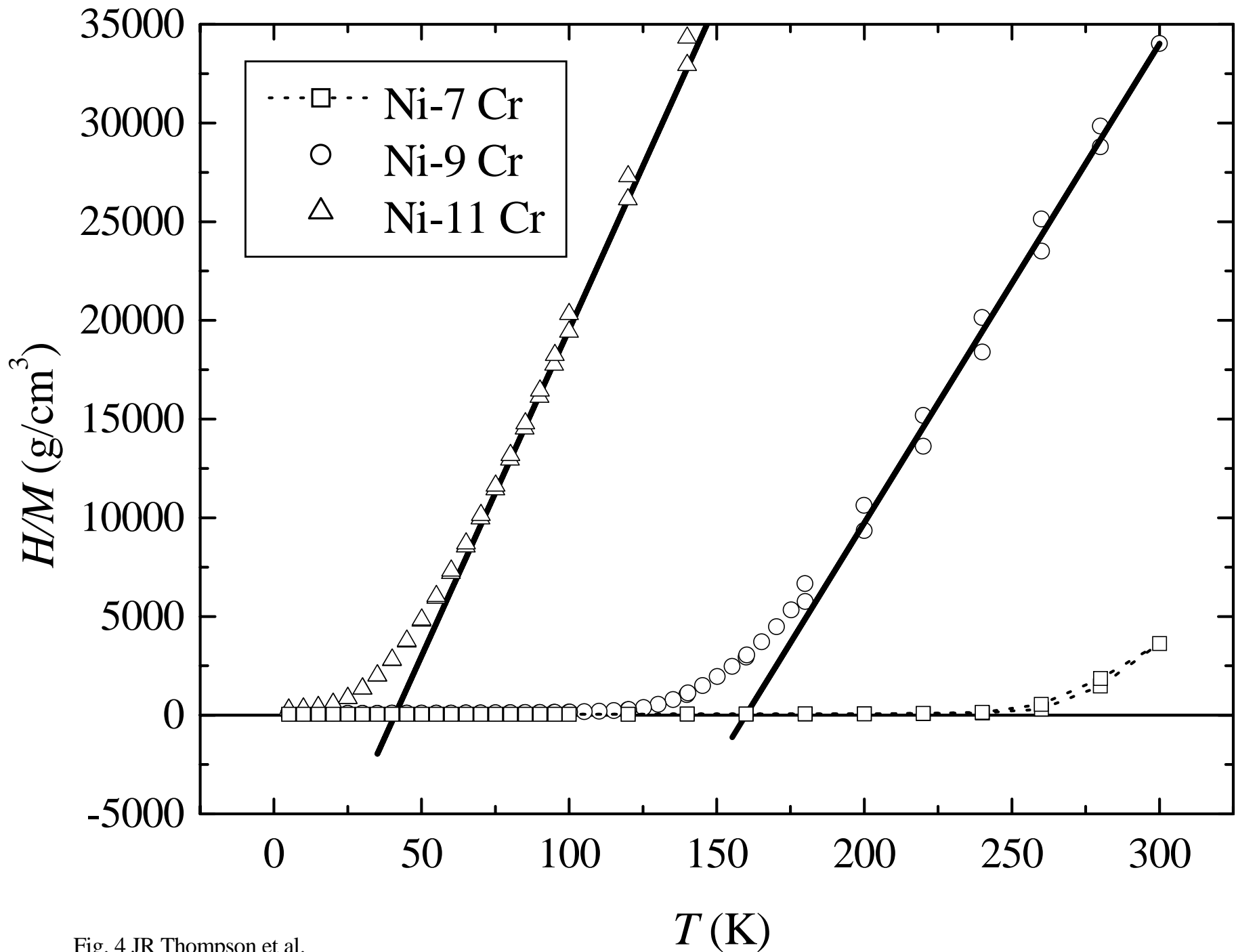

Fig. 4 JR Thompson et al.

C data NiCr

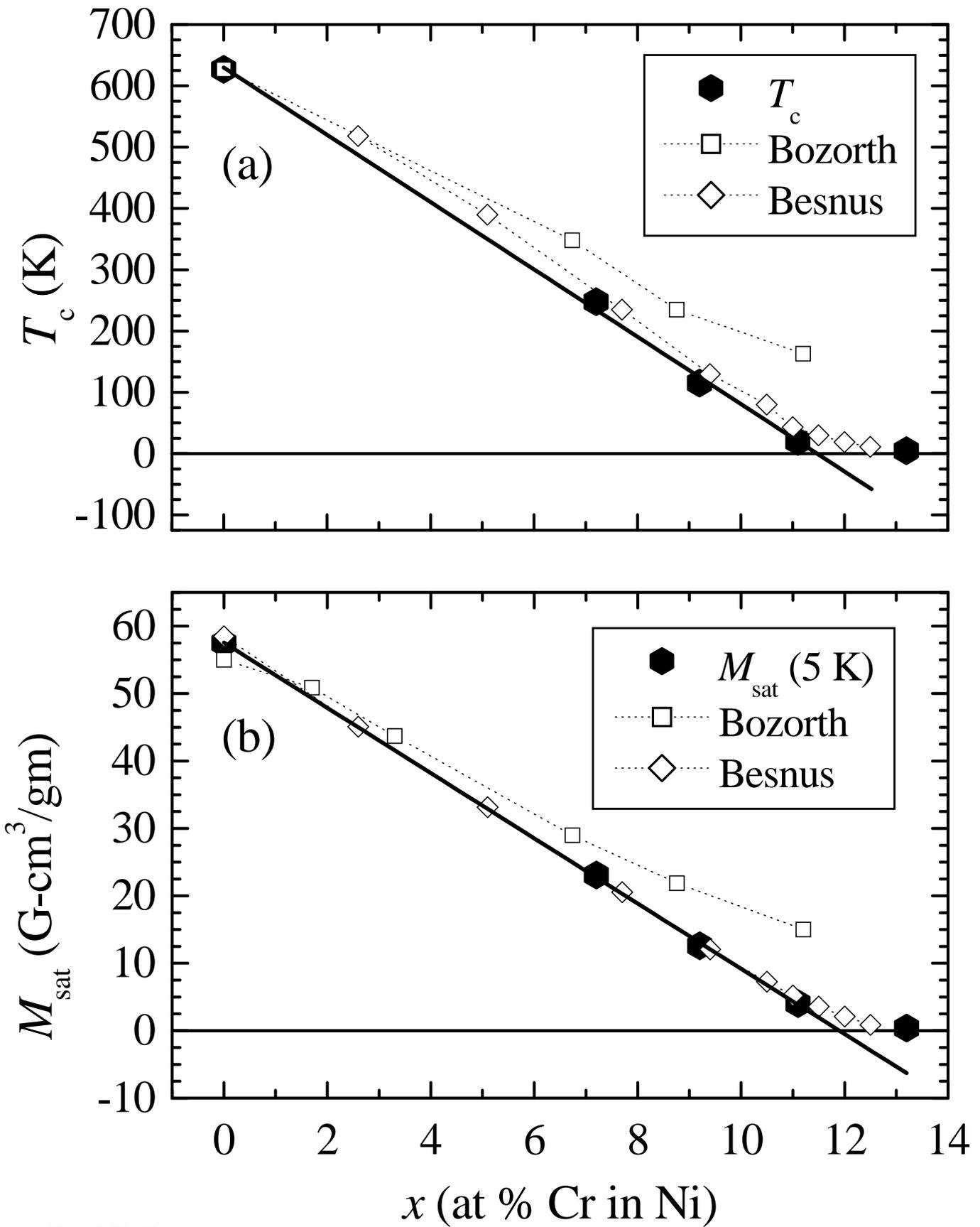

Fig. 5 JR Thompson et al.                                         C data NiCr

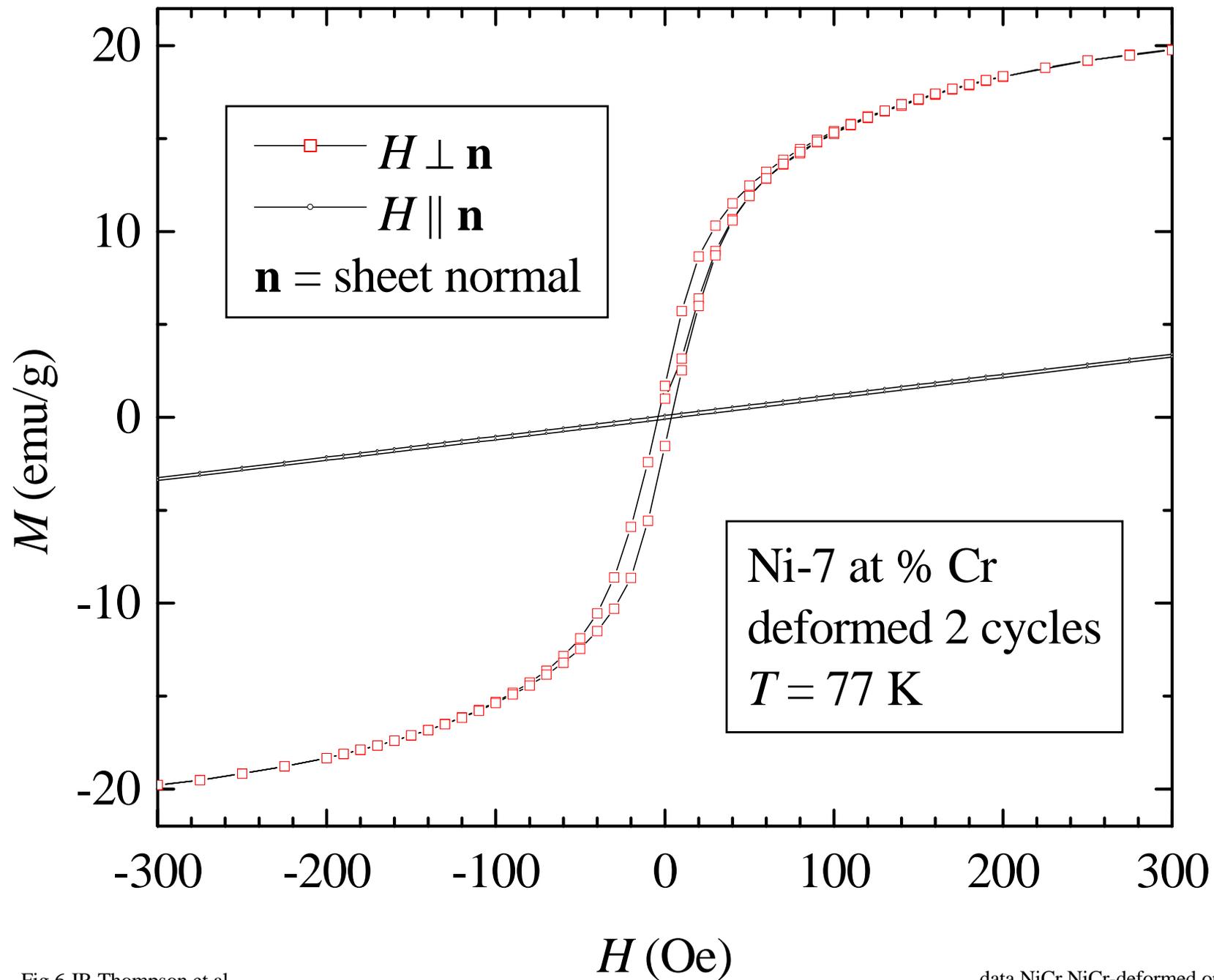

Fig 6 JR Thompson et al

data NiCr NiCr-deformed.opj

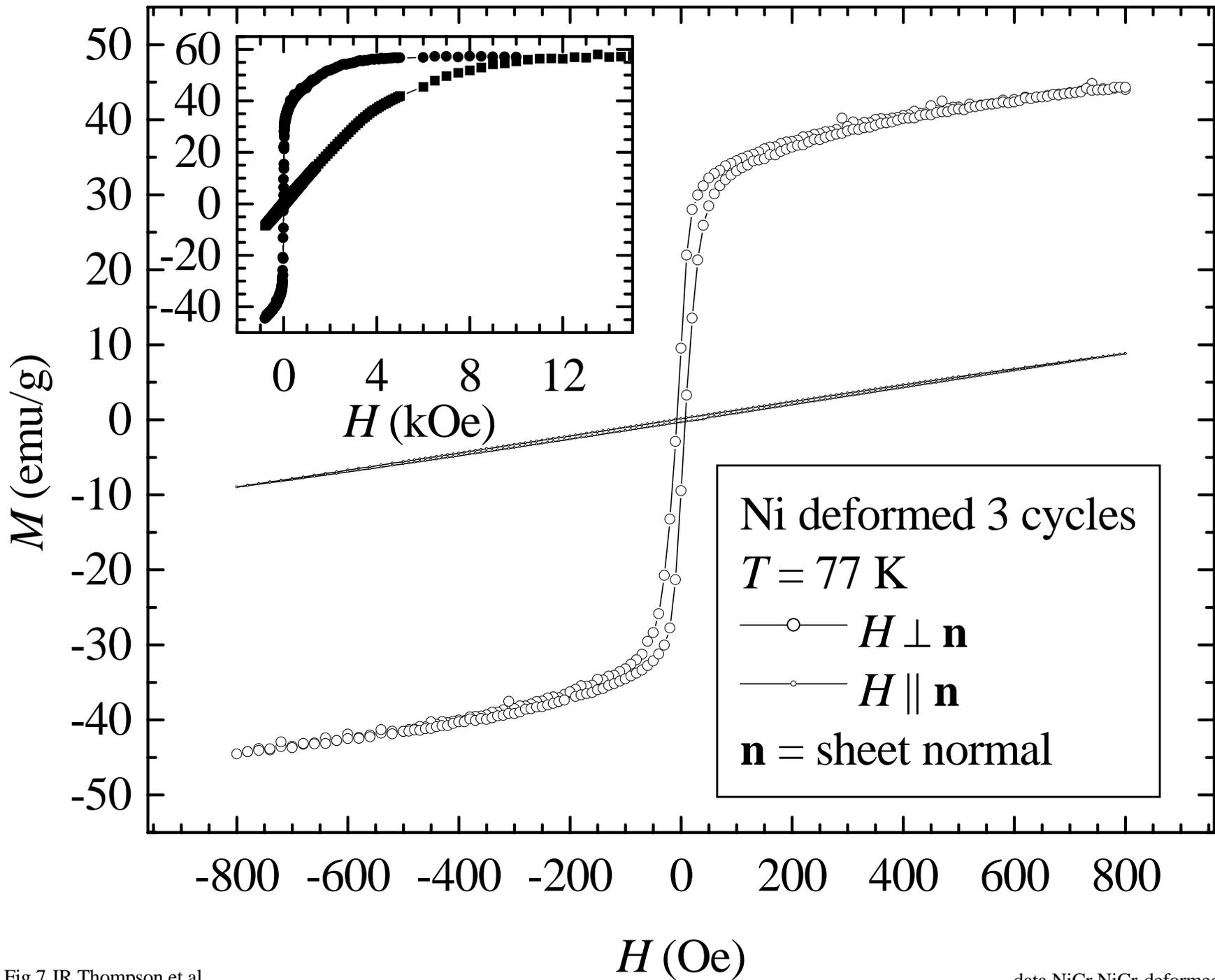

Fig 7 JR Thompson et al

data NiCr NiCr-deformed.opj

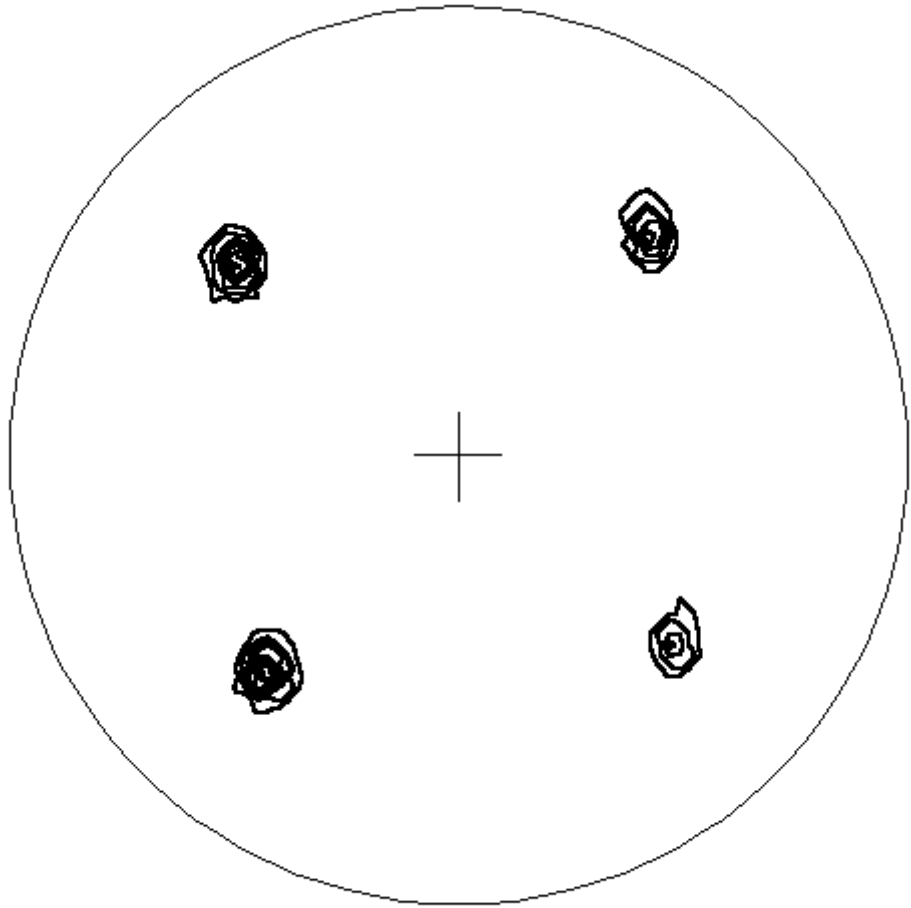

Fig. 8 JR Thompson et al.

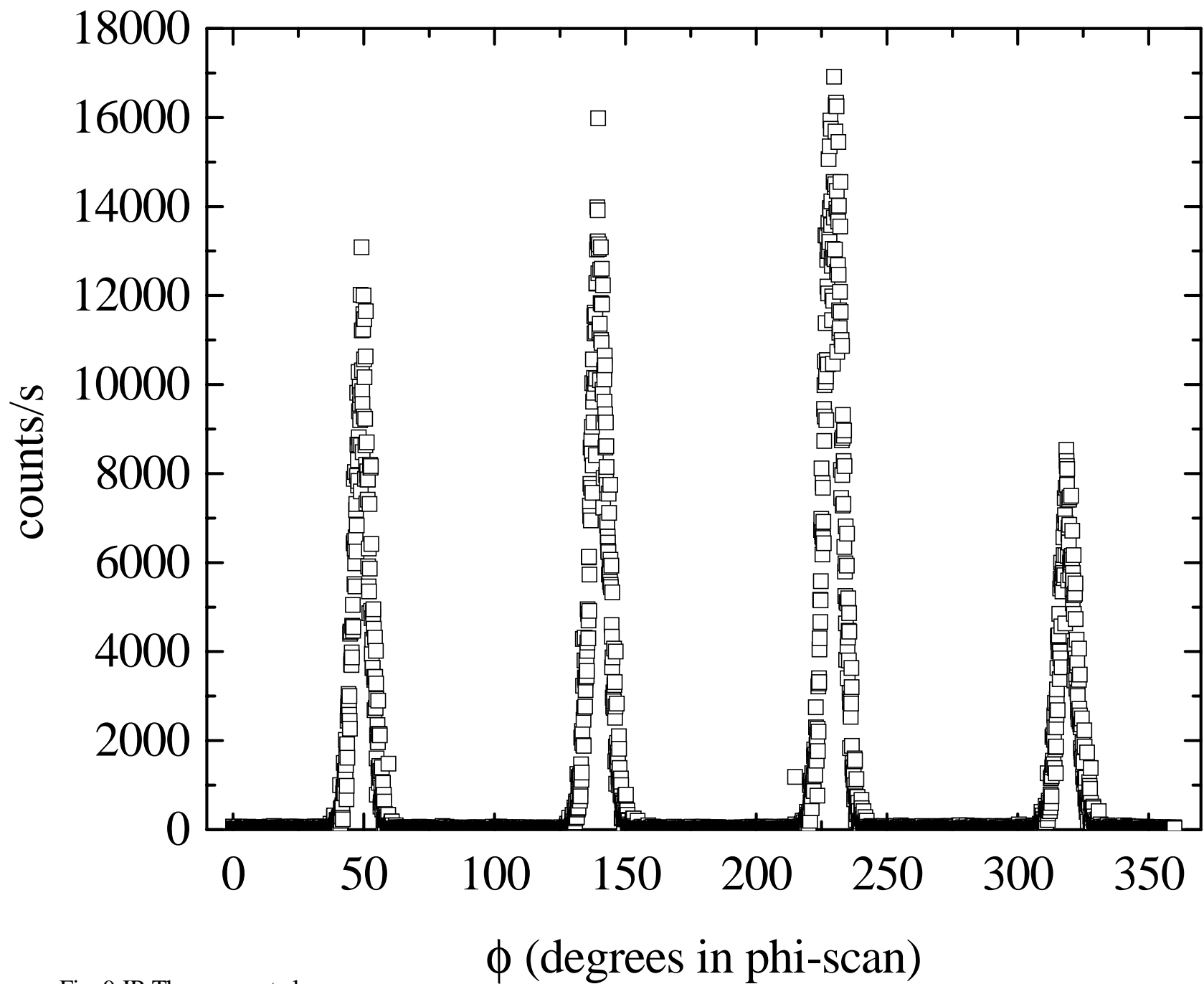

Fig. 9 JR Thompson et al.

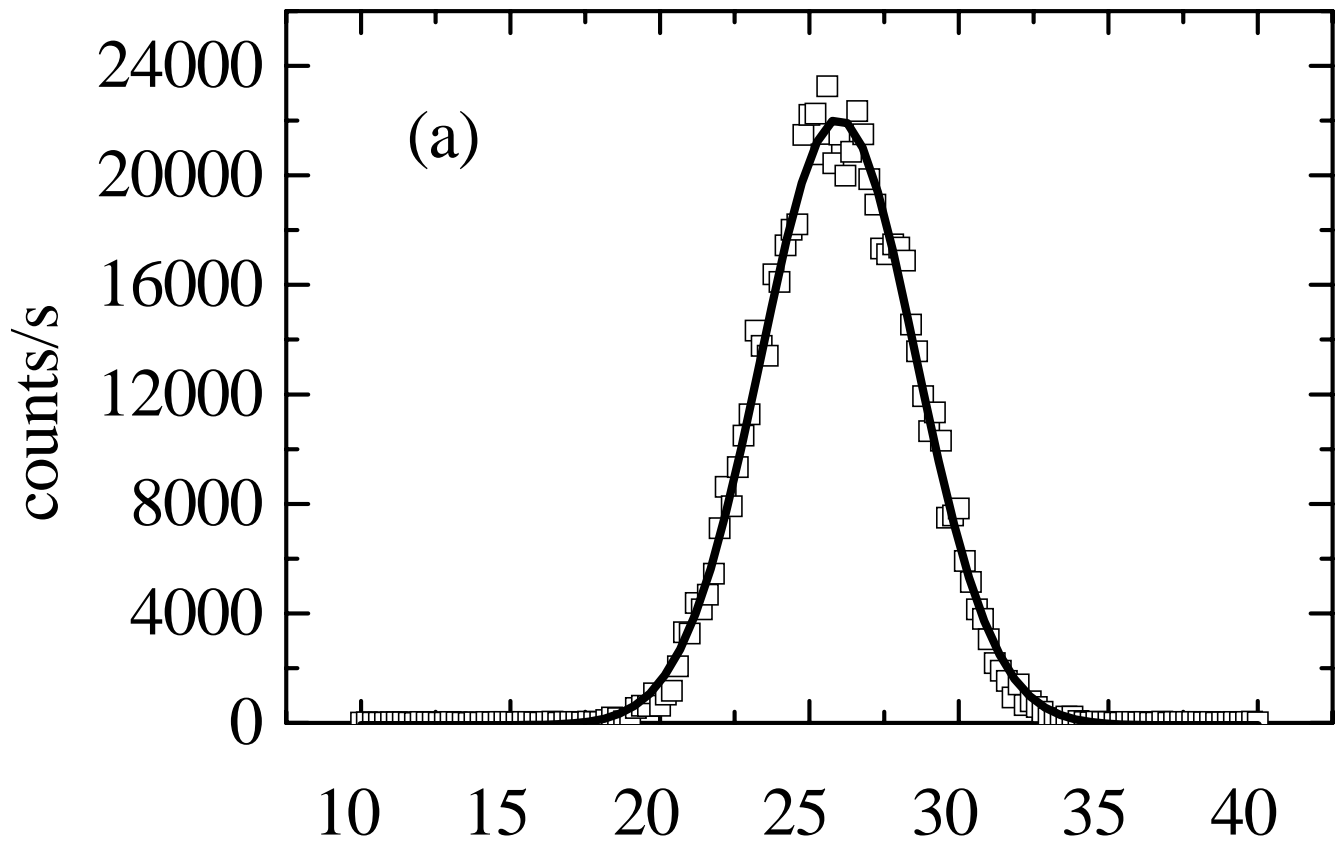

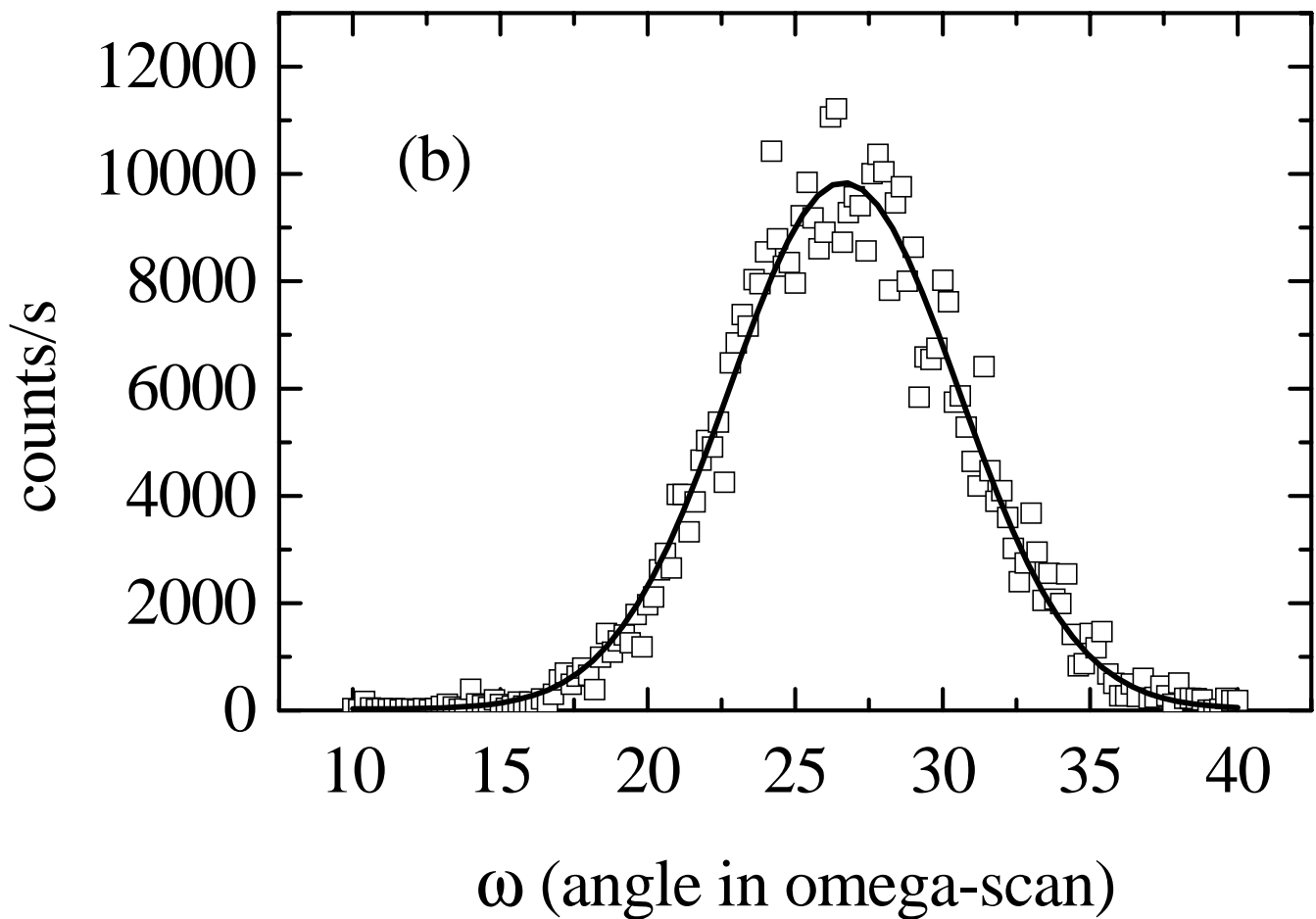

Fig. 10 JR Thompson et al.